# PrVO$_4$ under High Pressure: Effects on Structural, Optical and Electrical Properties

*Enrico Bandiello*[1,*], *Catalin Popescu*[2,*], *Estelina Lora da Silva*[3,4], *Juan Ángel Sans*[4], *Daniel Errandonea*[1], *Marco Bettinelli*[5]

[1]Departamento de Física Aplicada-ICMUV, MALTA Consolider Team, Universidad de Valencia, Edificio de Investigación, C/Dr. Moliner 50, Burjassot, 46100 Valencia, Spain

[2]CELLS-ALBA Synchrotron Light Facility, Cerdanyola del Valles, 08290 Barcelona, Spain

[3]IFIMUP, Departamento de Física e Astronomia, Faculdade de Ciencias da Universidade do Porto, Porto, Portugal

[4]Instituto de Diseño para la Fabricación y Producción Automatizada, MALTA Consolider Team, Universitat Politècnica de València, 46022 València, Spain

[5]Luminescent Materials Laboratory, Department of Biotechnology, University of Verona and INSTM, UdR Verona, Strada Le Grazie 15, 37134 Verona, Italy

[*]Corresponding authors:

Enrico Bandiello; email: enrico.bandiello@uv.es.

Catalin Popescu; email: cpopescu@cells.es

**Abstract**

In pursue of a systematic characterization of rare-earth vanadates under compression, in this work we present a multifaceted study of the phase behavior of zircon-type orthovanadate PrVO$_4$ under high pressure conditions, up until 24 GPa. We have found that PrVO$_4$ undergoes a zircon to monazite transition at around 6 GPa, confirming previous results found by Raman experiments. A second transition takes place above 14 GPa, to a BaWO$_4$-II -type structure. The zircon to monazite structural sequence is an irreversible first-order transition, accompanied by a volume collapse of about 9.6%. Monazite phase is thus a metastable polymorph of PrVO$_4$. The monazite-BaWO$_4$-II transition is found to be reversible instead and occurs with a similar volume change. Here we report and discuss the axial and bulk compressibility of all phases. We also compare our results with those for other rare-earth orthovanadates. Finally, by means of optical-absorption experiments and resistivity measurements we determined the effect of pressure on the electronic properties of PrVO$_4$. We found that the zircon-monazite transition produces a collapse of the band gap and an abrupt decrease of the resistivity. The physical reasons for this behavior are discussed. Density-functional-theory simulations support our conclusions.

## 1. Introduction

Due primarily to their technological and theoretical relevance, $RE$VO$_4$ rare-earth orthovanadates (with $RE$=trivalent rare-earth atom) are currently receiving a great deal of attention. In fact, as a consequence of their optical and luminescent properties, many of these materials are suitable for real-world applications, such as photocatalysts for the elimination of some organic pollutants and dyes,[1–3] as host materials for laser applications, luminescent emitters, thermophosphors and non-linear optics.[4–10]

These compounds are also challenging from the point of view theory and basic research, especially in the field of High-Pressure Physics. Sharing the same structure, in first approximation all zircon $RE$VO$_4$ have similar electronic and vibrational properties and follow a well-defined structural pathway under high-pressure conditions. They can be thus taken advantage of as a benchmark for theoretical simulations aimed at the prediction of their physical properties at ambient pressure and under high-pressure conditions. Moreover, the study of the high-pressure behavior of $RE$VO$_4$ may lead to the discovery of non-reversible transitions (i.e. metastable polymorphs), resulting in phases with useful physical properties. For instance, zircon NdVO$_4$ and SmVO$_4$ both suffer an irreversible phase transition to structures with a lower band gap than the zircon polymorph, a fact that may increase their efficiency in photocatalytic applications.[11,12]

In principle, NdVO$_4$ and compounds with larger $RE$ cations transform to a monoclinic monazite phase under pressure, while SmVO$_4$ and compounds with smaller cations experiment a transition to a scheelite structure.[13–15] This general trend can be altered by experimental conditions, as it has been established primarily using the most common quasi-hydrostatic pressure transmitting media (PTMs), such as silicon oil, methanol:ethanol or methanol:ethanol:water (ME, MEW).[16] As we will show in the following examples, alternative PTM choices may significantly alter the behavior of $RE$VO$_4$. For instance, a recent work of Marqueño et al. confirmed that NdVO$_4$ follows a zircon-scheelite-fergusonite transition sequence upon hydrostatic compression in Helium (He).[17]

For HoVO$_4$ the transition pressure from zircon to scheelite is almost halved when compressing it in a non-hydrostatic PTM. Moreover, a second phase transition, which is not observed in quasi-hydrostatic conditions, is found above 20 GPa.[18]

As a further example, the behavior of cerium vanadate CeVO$_4$ is somewhat erratic, showing evidence of an unstable equilibrium. This compound indifferently transforms from zircon to a scheelite or a monazite structure in a non-hydrostatic medium. On the contrary, it shows a well-defined zircon-monazite transition when compressed in a highly hydrostatic PTM.[19–21]

Finally, mixed vanadates such as Sm$_{0.5}$Nd$_{0.5}$VO$_4$ have an even more peculiar behavior, as after the first phase transition two different phases coexist (scheelite and monazite). In first analysis this could be attributed to the presence of crystalline domains with excess of Nd or Sm. Nonetheless, the homogeneity of the samples and a comparison with the behavior isostructural TmPO$_4$ phosphate led to the conclusion that this effect is instead due to a very small difference in enthalpy between the coexisting phases.[22,23] All these examples show the rich phenomenology exposed by the systematic investigation of RE vanadates and confirm the need for further insights in this area.

Here we present a comprehensive study of zircon-type $PrVO_4$ under high pressure conditions, up to about 24 GPa, by means of powder X-ray diffraction, UV-VIS absorption spectroscopy and electrical resistivity measurements. Similarly to $SmVO_4$, this compound is investigated for its use as a photo-degradant for organic dyes.[24] On the other hand, high-pressure studies on $PrVO_4$ under high pressure-conditions are scarce. To the best of our knowledge, the only previous work on this topic is the one by Errandonea et al., which reported a transition from zircon to a monazite phase by means of Raman spectroscopy.[12] In this work, we have observed two different phase transitions for $PrVO_4$ in the range 1 bar – 24 GPa. Using powder samples and synchrotron radiation, in first instance we have confirmed the results obtained by Errandonea et al.,[12] i.e. a zircon-monazite transition is found above 6 GPa. This transition implies an atomic rearrangement and a coordination change of the $Pr^{3+}$ ion from 8 to 9, together with a large volume collapse, of about 9.6%. Subsequently, we have found a further phase transition around 14.4 GPa, from monazite to a $BaWO_4$-II -type structure, which belongs to the same space group of monazite but has 8 formula units per unit cell, instead of 4. Furthermore, optical-absorption and resistivity measurement show drastic changes induced by pressure in the electronic properties of $PrVO_4$, which can be associated to the zircon-monazite transition. Our findings are supported by density-functional-theory (DFT) calculations and are discussed here in the general framework of lanthanide orthovanadates under compression.

## 2. Experimental Details

$PrVO_4$ single crystals were synthetized by the flux method, following a process that has been described elsewhere in detail for analogous $REVO_4$ orthovanadates.[11] The obtained samples were optically transparent and up to 1x1x10 mm$^3$ in size.

High-pressure angle-dispersive powder X-ray diffraction (P-XRD) experiments were performed at room temperature (RT) employing a membrane diamond-anvil cell (DAC), with diamond culets of 400 μm. For our experiments, selected crystals with no flux inclusions were chosen and carefully grinded down to a fine powder. The powder was loaded in a 180 μm hole drilled on a stainless-steel T301 gasket pre-indented to a thickness of 40 μm. Copper powder (Cu) was used as the pressure gauge and a methanol:ethanol:water 16:3:1 (MEW) mixture was used as the pressure-transmitting medium (PTM). Experiments were performed at the BL04-MSPD of ALBA Synchrotron with a monochromatic X-ray beam (λ = 0.4246 Å) focused to a 20 μm × 20 μm spot (FWMH).[25] XRD patterns were collected with a Rayonix CCD detector. After an initial mapping of the whole sample, the X-ray spot has been kept in the position which offered the best signal-to-noise ratio. Structural analyses were performed with GSAS-II and MAUD, using the Le Bail method.[26,27] Visualization of crystal structures has been done with VESTA.[28] EosFit7-GUI has been used to calculate the bulk modulus by fitting the pressure-volume data with the Birch-Murnaghan equation of state (EOS).[29]

For optical absorption measurements we used 10-μm thick platelets cleaved from the single-crystals. The experiments were performed in a membrane-type DAC with diamond culets of 500-μm. The pressure medium was 4:1 ME and the ruby scale was used to determine the pressure.[30] Absorption spectra were measured by comparing, at each pressure, the spectrum of the light source through the pressure transmitting medium and through the sample ("sample-in, sample out method"). Resistivity measurements up to 10 GPa were carried out using an opposed Bridgman-anvil setup.[31] The pressure was determined by calibrating the hydraulic pressure of the press against known phase transitions.[32] A sintered powdered sample, made from the sample used in Ref. 12, with 3 mm × 3 mm × 0.1 mm dimensions was used in this experiment. The measurements were carried out in a four-point configuration.

### 3. Theoretical Framework

To compute the structural and electronic properties of the three different crystalline phases of $PrVO_4$, calculations were carried out within the framework of density-functional theory (DFT).[33] The Vienna *ab-initio* Simulation Package (VASP)[34] was employed to perform calculations with the projector augmented-wave (PAW) scheme. Electronic convergence was obtained with a plane-wave kinetic-energy cut-off of 700 eV. The Brillouin-zone (BZ) was sampled with Γ-centered Monkhorst-Pack[35] grids by employing the following subdivisions for the different structural phases for the three compounds: zircon and monazite with 12$x$12$x$12, and *post-monazite (BaWO$_4$-II)* with 6$x$12$x$12. The generalized-gradient approximation (GGA)[36] with the Perdew, Burke, and Ernzerhof (PBE) parameterization was applied for all the calculations, together with the DFT+$U$ method proposed by Dudarev *et al.*,[37] to enable the treatment the strongly correlated *d*-states of V. The Dudarev method employs the rotationally invariant approach,[37] where the correction for the on-site interactions between the V *d*-electrons is expressed by a single $U_{eff}$ parameter, where $U_{eff}=U-J$, with $U$ and $J$ being the screened Coulomb and the exchange parameters, respectively. The value for $U_{eff}$ was chosen to be 3.25 eV, which is an adequate value in order to treat the localized d-states of V systems (Refs. [38] and [39] for more details). The lower states of the conduction band are mainly constituted by the V *d*-states whereas the O-*p* constitutes the upper valence band which were treated at the PBE level. By considering existing literature,[39] the Pr *f*-states are treated at the core level ("frozen" inside the core region). Since Pr is trivalent ($Pr^{3+}$) there is no need to consider the empty states of Pr. Also, and since the valence band maximum (VBM) and conduction band minimum (CBM) are formed mainly by the V *3d*- and O *2p*-states, respectively, the *f*-states of Pr will have no effect on the bonding environment. The Hubbard $U_{eff}$ potential is hence only required to treat the localized *d*-states of the V system, whereas all other states are treated at the PBE level.

The atomic positions and the unit-cell parameters were fully optimized to obtain the relaxed structures at selected volumes. From the relaxed optimized configurations, the resulting forces on the atoms were less than 0.001 eV/Å. The calculations provide not only a set of structural parameters as a function of pressure, but also a set of accurate energy, volume and pressure data that are fitted to the third-order Birch-Murnaghan equation of state (BM EoS) in order to obtain the equilibrium volume, the bulk modulus and pressure derivatives.[40,41]

The electronic band structures were computed by including spin-orbit coupling (SoC) effects due to the heavy elements that constitutes the system, namely due to Pr, and the post-processing was carried out by employing the SUMO python toolkit.[42]

Enthalpy calculations, *H*, are carried out to probe the thermodynamical stability of the system and is obtained by fitting the sum of total energy, *E*, and the pressure and volume, $P\times V$, terms to a 4$^{th}$ order polynomial ($H=E+P\times V$). From the current fit it is possible to obtain the relative *H* with respect to the pressure.

## 4. Results and Discussion

As most $RE$VO$_4$ vanadates (except LaVO$_4$), PrVO$_4$ crystallizes naturally in a zircon structure (S.G. 141, $I4_1/amd$, $Z$=4), in which the vanadium atom is surrounded by four oxygen atoms in a tetrahedral coordination and the trivalent Pr cation is eight-fold coordinated with oxygen atoms, forming PrO$_8$ bidisphenoids (Figure 1). Experimental unit-cell parameters for zircon PrVO$_4$ at ambient conditions, as measured by powder XRD, are $a$ = 7.3625(2) Å, $c$ = 6.4614(4) Å, with a volume $V$ = 350.3(5) Å$^3$. These values are in good agreement with previous reports, from which they differ by less than 0.1% (Table 1) and with *ab initio* DFT calculations (Table1).[43] The theoretical parameters of the unit cell of the zircon phase are slightly overestimated when compared to the experimental data, which is expected for GGA due to the well-known overestimation of the equilibrium volumes.[44] Moreover, the inclusion of the Hubbard $U$ parameter will also affect the lattice parameters of the unit-cell. However, the volume is in line with the data found in the literature.[39] As we will observe in the following, the overestimation of the volume and the lattice parameters holds true for all the phases of PrVO$_4$ and in all the pressure range.

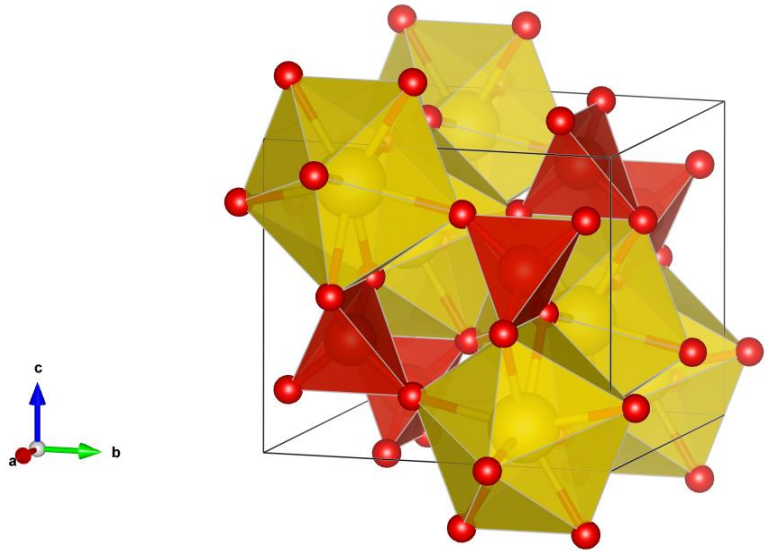

**Figure 1.** Zircon structure of PrVO$_4$. The coordination polyhedra of Pr and V are in yellow and red, respectively.

| | | zircon PrVO$_4$ | | |
|---|---|---|---|---|
| | | *cell parameters* | | |
| | | $a$ (Å) | $c$ (Å) | $V$ (Å$^3$) |
| Experiment | | 7.3625(2) | 6.4614(4) | 350.3(5) |
| DFT | | 7.46851 | 6.53544 | 364.54 |
| Experiment ( Ref. [43]) | | 7.3631(1) | 6.4650(1) | 350.5(7) |
| | | *atomic positions* | | |
| *atom* | *Wyckoff position* | *x/a* | *y/b* | *z/c* |
| Pr | 4a | 0 | 1/4 | 3/8 |
| V | 4b | 0 | 3/4 | 1/8 |
| O | 4c | 0 | 0.42944 (*0.4288(1)*) | 0.70352 (*0.7057(1)*) |

**Table 1.** Experimental and theoretical cell parameters for zircon PrVO$_4$, as obtained in this work, and comparison with the literature (Ref. 43). Experimental atomic positions from Ref. 43 (in parenthesis) are shifted by (0, 0,+0.5) for easier comparison.

Simulated atomic positions are also shown in Table 1 and are in agreement within a 0.1% interval with previous experimental results.[43] In Figure 2a the powder XRD patterns for PrVO$_4$ at selected pressures are shown. Patterns below 6.2 GPa are correctly indexed by the zircon structure and show mild changes with pressure, such as the usual shift towards higher values of 2θ due to the shrinking unit cell. On the other hand, at 6.2 GPa sudden changes appear, because of the onset of a phase transition. The new phase can be indexed in the monazite structure (S.G. 14, $P2_1/n$, Z=4), with unit-cell parameters $a$ = 6.9930(8) Å, $b$ = 7.1796(7) Å, $c$ = 6.6311(8) Å and $\beta$ = 104.60(1)° ($V$ = 322.2(8) Å$^3$). As anticipated, DFT calculations predict similar values for the cell parameters, although the cell volume is slightly overestimated (See Table S3). This structure is represented in Figure 3a, and it consists of PrO$_9$ polyhedra and VO$_4$ tetrahedra, with an increase in the coordination of Pr. As the monazite phase is stable upon pressure release, the cell parameters have been obtained at 0.4 GPa, the residual pressure of our DAC. These values are found to be fully consistent with previous results obtained for PrVO$_4$.[12] The range of stability of the monazite phase is limited, as new features arise in the XRD patterns just above 6.8 GPa. The changes in the patterns are highlighted in Figure 2b and they indicate the onset of a new transition to a post-monazite phase. As we can observe in Figure 2a, the Bragg peaks become broader after the first transition, probably due to strain in the crystal grains and to a large volume collapse (more details below). This is evident from the 2D diffraction images above 6.2 GPa, all of which show very broad and diffuse lines in correspondence of the Bragg peaks of PrVO$_4$. Also, the texture of the sample is not significantly affected by the transition (Figure 2c). The quality of our XRD patterns above 6.2 GPa does not allow thus for a full Rietveld refinement. Upon comparison with literature, we found that a sensible candidate for the post-monazite phase is the monoclinic BaWO$_4$-II structure (S.G. 14, $P2_1/n$, Z=8), which belongs to the same space group of monazite with a unit cell twice as large (Figure 3b). An analogous transition has in fact been reported before for monoclinic LaVO$_4$.[45] The BaWO$_4$-II structure is able to explain all the peaks of the post-monazite phase, as shown in Figure 4, where some remnants of the monazite phase are still visible. Above 10 GPa, the diffraction peaks are broadened, due to the deviatoric stresses introduced by the loss of hydrostaticity of the PTM.[16,46] Reliable cell parameters for the BaWO$_4$-II phase could only be obtained at higher pressure, at around 14.4 GPa (see Figure 4), and are $a$ = 11.557(7) Å, $b$ = 6.522(7) Å, $c$ = 6.847(7) Å and $\beta$ = 90.71(1)° ($V$ = 516.1(8) Å$^3$). The DFT calculated post-monazite phase (at 14.93 GPa) presents a qualitative agreement with the structure proposed from experiment (Table S3), for which the calculated $a$ parameter is found to be somewhat higher than the experimental value. This accounts for the overestimation of the equilibrium volume (Table S3). It is also interesting to note that, unlike what happens for the monazite phase, the theoretical $\beta$ angle for the BaWO$_4$-II phase is also overestimated (96.2° vs. 90.7°). One reasonable explanation for this discrepancy is the distortion of the BaWO$_4$-II structure due to the non-hydrostaticity of the pressure medium above 10 GPa.[16]

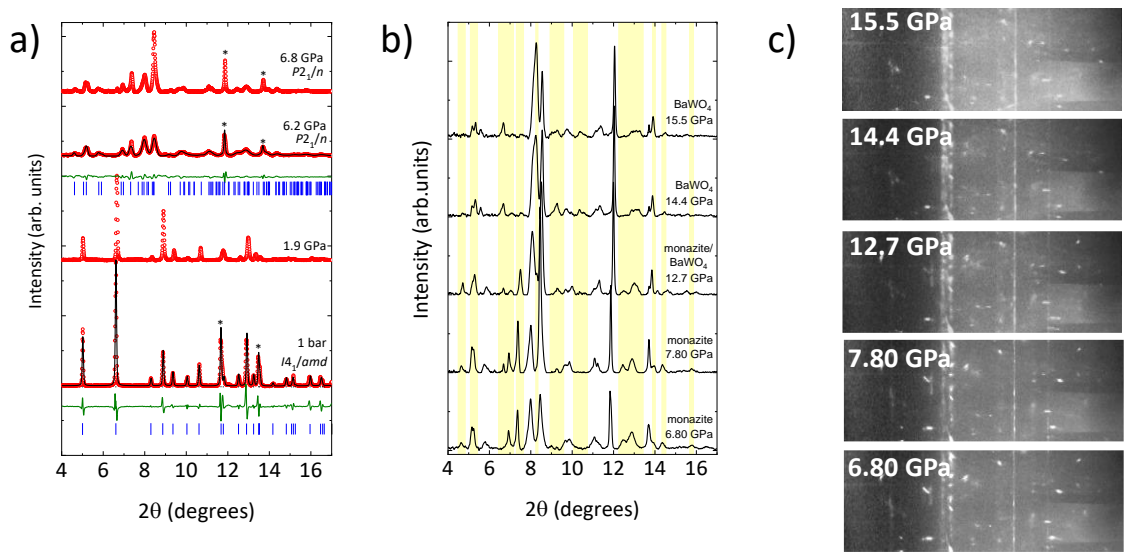

**Figure 2.** a) XRD patterns for PrVO$_4$ at selected pressures (open red circles). The pressure $P$ is indicated in GPa above each pattern; Le Bail refinements (black lines) are shown for the zircon phase at ambient pressure and the monazite phase at 6.2 GPa. Green lines: residuals. Blue ticks: Bragg reflections; b) XRD patterns from 6.8 to 15.5 GPa showing the zones with the most prominent changes during the phase transition from monazite to BaWO$_4$-II highlighted in yellow; the majority phase is indicated above each pattern; c) unrolled ana non-masked CCD cake images, for 2θ≈4º - 17º, corresponding to the patterns in Figure 2b. Cu reflections are also visible. Spots whose position does not change upon compression are not attributable to the sample under examination have been masked out in the XRD analysis.

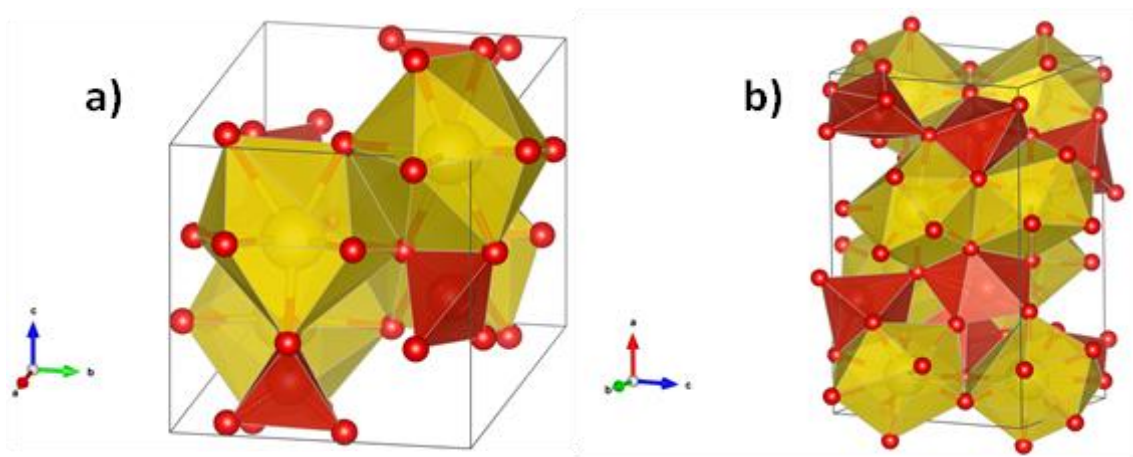

**Figure 3.** Unit cell of a) monazite and b) BaWO$_4$-II phase of PrVO$_4$. The coordination polyhedra of Pr and V are shown in yellow and red, respectively. Small red spheres are oxygens. Atomic positions for monazite were taken from Ref. 12.

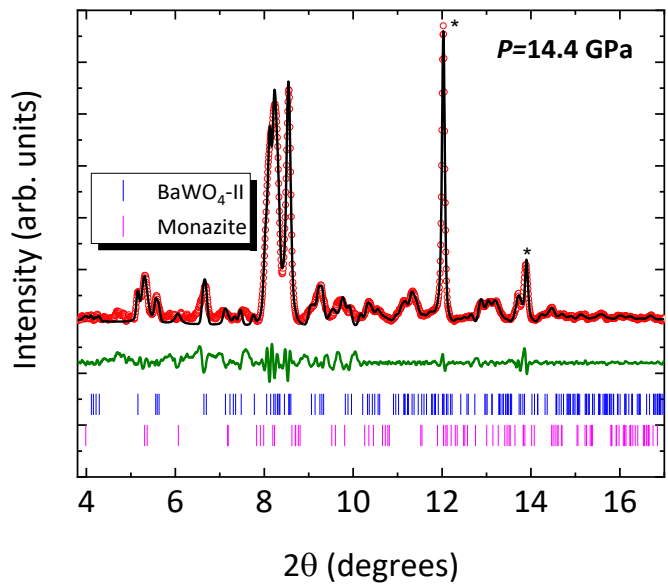

**Figure 4.** XRD patterns for PrVO$_4$ at 14.4 GPa (open red circles). Black lines: Le Bail refinement; green lines: residuals; magenta (resp. blue) ticks: Bragg reflections for the monazite (resp. BaWO$_4$-II) structure. Asterisks are shown in proximity of the Cu peaks.

BaWO$_4$-II phase is stable up to the maximum pressure reached in our experiment, which is around 23.4 GPa. Upon pressure release the system recovers the monazite phase, which is found to be stable even close to ambient pressure (Figure 5). As shown also in previous reports, monazite PrVO$_4$ is thus confirmed to be a metastable polymorph of PrVO$_4$.[12] The phase-transition sequence observed here is fully supported by DFT calculations. We have plotted the enthalpy *vs* pressure curves to probe the thermodynamic stability of the three phases of PrVO$_4$ (Figure 6). Zircon is the most energetically stable phase at room pressure, as experimentally observed. At ambient pressure the monazite and the BaWO$_4$-II phase are 0.04 and 0.13 eV higher in enthalpy than zircon, respectively. For increasing pressure, the monazite system becomes energetically more stable, with a transition occurring from zircon to monazite at around 6.02 GPa, consistently with our experimental results. Upon further compression, the BaWO$_4$-II structure competes with monazite and at around 11.11 GPa it becomes energetically more favorable. Again, this agrees well the results obtained from our experiments. The enthalpy difference between monazite and BaWO$_4$-II remains within 0.01 eV around the crossing point, which may explain the phase coexistence observed in the 11 – 14 GPa pressure range. Finally, the BaWO$_4$-II phase appears to be the most thermodynamically stable among the three structures considered in the present work, up to the maximum pressure reached in our simulations (~25 GPa).

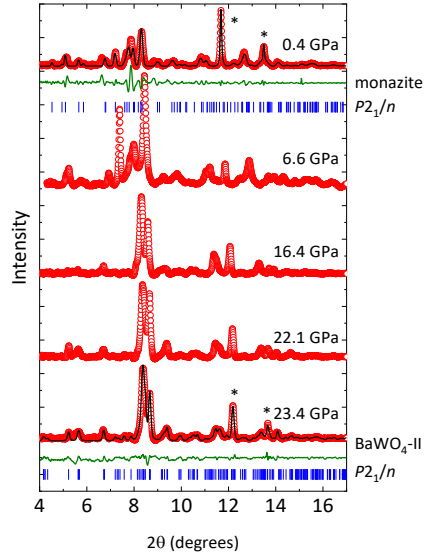

**Figure 5.** XRD patterns at pressure release, showing the gradual recovery of the monazite phase form BaWO4-II structure. The pressure $P$ is indicated in GPa above each pattern; Le Bail refinements (black lines) are shown for the BaWO$_4$-II-type phase at 23.4 GPa and for the monazite phase at pressure release, 0.4 GPa. Green lines: residuals. Blue ticks: Bragg reflections. Asterisks are shown in proximity of the Cu peaks.

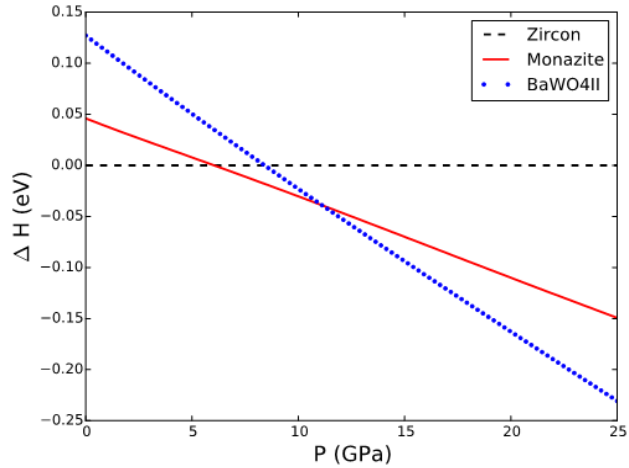

**Figure 6.** Relative enthalpy differences of the three different phases of PrVO$_4$, with respect to the enthalpy of the zircon structure.

Now we will discuss the axial compressibility of each one of these structures. This physical quantity is defined as $k_s = -\frac{1}{s_0}\frac{\partial s}{\partial P}$, where $s$ is one of the cell parameters ($a$, $b$ or $c$), $s_0$ is the perspective value at ambient pressure and $P$ is the pressure. The values of the compressibility of each axis for the zircon phase are reported in Table S1 (Sup. Material) and were obtained by performing linear fits on the data points of Figure 7. Notice that the unit-cell parameters obtained for monazite-type PrVO$_4$ agree well with those obtained from quenched samples in a large volume press (open diamonds in Figure 7).[12]

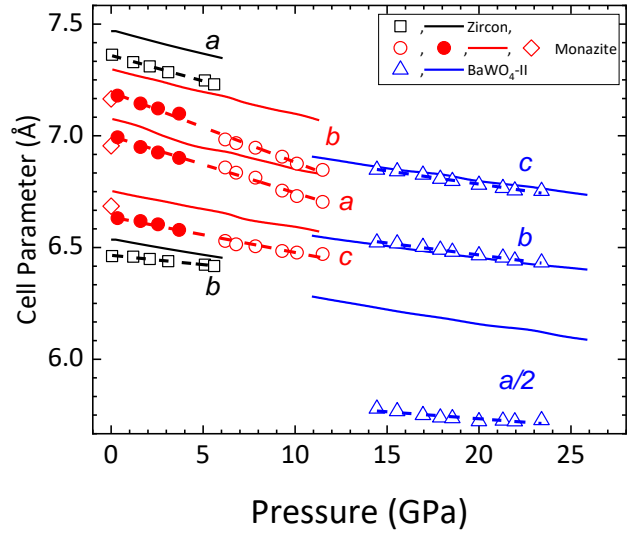

**Figure 7.** Symbols: Experimental cell parameters of the zircon (black squares), the monazite (red circles) and the BaWO$_4$-II phases (blue triangles) as a function of pressure. Dashed lines: linear fits of the cell parameters (black: zircon; red: monazite; blue: BaWO$_4$-II phase). Continuous lines: data calculated by *ab initio* simulations (black: zircon; red: monazite; blue: BaWO$_4$-II phase). Open diamonds are the parameters of the monazite phase from Ref. 12. Closed black circles are experimental data for monazite taken during pressure release.

The axial compressibility of the zircon phase is strongly anisotropic (Figure 7), as it is larger along *a* than along *c*, with $k_a = 3.1(1) \times 10^{-3}$ GPa$^{-1}$, $k_c = 1.29(8) \times 10^{-3}$ GPa$^{-1}$. The anisotropy of the zircon phase is not unusual for *RE*VO$_4$ systems, due to their polyhedral arrangement, and in the case of PrVO$_4$ it is in line with previous reports.[13,17,47] For zircon PrVO$_4$ the volume of the VO$_4$ units is almost 11 times lower than that of the PrO$_8$ polyhedra, implying that at first order the compressibility is dominated by the changes of PrO$_8$ molecular units. For this zircon phase, the PrO$_8$ units lie in zig-zag chains along [100] and [010] and are stacked along [001], separated by the scarcely compressible VO$_4$ tetrahedra (Figure 1). This explains why zircon PrVO$_4$ is preferentially compressed along [100] (and [010]) rather than along [001]. The analysis of the axial compressibility of the monazite and post-monazite must be carried out by calculating the isothermal compressibility tensor for these phases. According to the explicit expressions reported by Knight, for monoclinic structures the compressibility tensor is symmetric and with only 4 non-zero components.[48] For PrVO$_4$, the tensor for monazite and post-monazite structures ($\beta_{ij}$) and corresponding eigenvectors and eigenvalues ($\lambda_i$ and ev$_i$, respectively) have been calculated using the aforementioned expressions and reported in Table S2, using both experimental and theoretical cell parameters as a function of pressure. For monoclinic structures one of the eigenvectors is always parallel to the monoclinic *b* axis.[48] Both monoclinic phases of PrVO4 present anisotropic compressibility along the principal axes. In both cases, the least compressible axis is orthogonal to the monoclinic *b* axis (parallel to eigenvector $ev_2$ in Table S2).[48]

For the three phases, the unit-cell length parameters are overestimated by the DFT simulations, in particular the $a$ axis of the BaWO$_4$-II phase. This is expected, given that DFT generally overestimates unit-cell volumes. Nonetheless, theoretical simulations correctly predict anisotropic compressibility for all the phases, in agreement with our experiments (Table S2). For completeness we also calculated the pressure coefficient of the $\beta$ angle, $k_\beta$, for both monoclinic structures: $\beta$ angle decreases with pressure for the monazite, while it has a slightly increasing trend for the BaWO$_4$-II phase ($k_\beta$ = -0.074(6) and 0.018(13) deg GPa$^{-1}$, respectively) (Figure 8). In addition, the $\beta$ angle of the monazite phase suffers a small discontinuity at the transition, after the pressure release. We believe that this is due to strain induced in the crystal grains after two subsequent phase transitions, which also causes the broadened peaks for monazite in Figure 2a. DFT simulations accurately predict the $\beta$ angle of the monazite phase at ambient pressure and its derivative (Figure 8). On the other hand, the calculated $\beta$ angle for the BaWO$_4$-II structure is consistently larger than the experimental value, as anticipated, and it is predicted to monotonically decrease with pressure, unlike to what has been observed during our experiments. As stated before, we believe that the reason behind these discrepancies is the loss of hydrostaticity of the PTM above 10 GPa, which may induce distortion in the BaWO$_4$-II unit cell. In our refinements we have assumed the atomic positions reported by Errandonea et al. for monazite PrVO$_4$,[12] therefore the monazite unit-cell presents the layout shown in Figure 3a. Also, being monazite stable close to ambient pressure, the cell parameters of monazite for $P < 6.2$ GPa have been considered in our calculations. Regarding the anisotropic compressibility of monazite, a qualitatively similar behavior has been observed for monazite-type LaVO$_4$,[14] monazite-type CeVO$_4$, and monazite-type phosphates.[49,50]

A reduced axial compressibility is found for the BaWO$_4$-II phase. Two of the principal axes have roughly the same compressibility while the third one is slightly stiffer. In addition, the $\beta$ angle of the BaWO$_4$-II structure is less dependent on pressure than the monoclinic angle of monazite. As we will see in the following, a further volume collapse takes place at the second transition, increasing the packing of the structure. This feature reduces the compressibility of the post monazite phase.

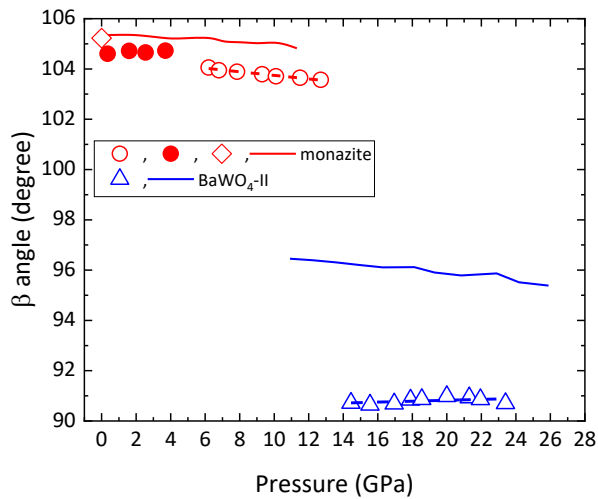

**Figure 8.** Experimental trend of the $\beta$ angle of monazite (red circles) and BaWO$_4$-II-type (blue triangles) as a function of pressure. Open circles for monazite are data taken during pressure release. The diamond

indicates the angle of monazite PrVO$_4$ at ambient pressure as reported in Ref. 12. Dashed lines are linear fit of the experimental data, while values calculated by DFT are shown as continuous lines (red: monazite; blue: BaWO$_4$-II phase).

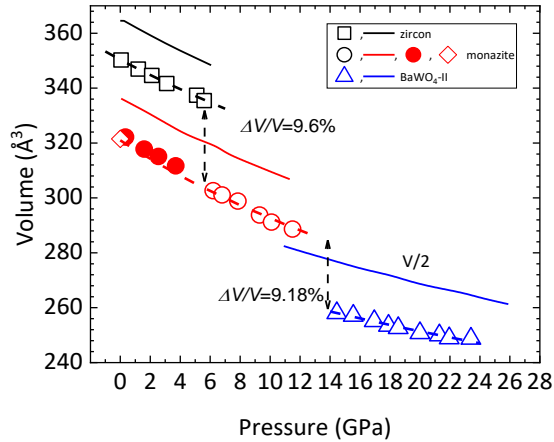

**Figure 9.** Volume as a function of pressure for the different phases of PrVO$_4$. Symbols: Experimental values of volume of the zircon (black squares), the monazite (red circles) and the BaWO$_4$-II phases (blue triangles). The open diamond is the volume of the monazite phase from Ref. 12. Closed red circles are experimental data for monazite taken during pressure release. Dashed lines: EOS fits of the experimental values (black: zircon; red: monazite; blue: BaWO$_4$-II phase). Continuous lines: data calculated by *ab initio* simulations (black: zircon; red: monazite; blue: BaWO$_4$-II phase).

In the following we will comment on the EOS of PrVO$_4$. Pressure-volume (*P-V*) data are shown in Figure 9. The bulk moduli of all polymorphs have been estimated by fits of the pressure-volume data with a 2$^{nd}$ order Birch-Murnaghan EOS (Table 2). The first transition has a first-order nature, since it is irreversible and a volume collapse of about 9.6%, which is measured as the difference at the EOSs, is detected at the onset of transition ($P = 6.2$ GPa). This is in agreement with previous reports for the zircon-scheelite transition for other rare-earth orthovanadates.[13] For zircon PrVO$_4$, the initial volume $V_0$ and the bulk modulus $B_0$ resulting from the fits are $V_0$=350.4(3) Å$^3$, $B_0$=120.3(3) GPa. The calculated ambient-pressure volume is in good agreement with the experimental value. Similar bulk moduli have been found in the isostructural CeVO$_4$, TbVO$_4$ and TmVO$_4$ vanadates.[21,47] For an easier comparison, EOSs of other *RE*VO$_4$ in zircon, monazite and BaWO$_4$-II are summarized in Table 2. Under compression in non-hydrostatic media, most zircon *RE*VO$_4$ vanadates show larger bulk moduli, often above 130 GPa.[13] Zircon PrVO$_4$, together with TmVO$_4$, and zircon-phase LaVO$_4$ is thus one of the softest vanadates (Table 2). This is true also for the monazite phase, which has a bulk modulus of 95(6) GPa. This is a very low value, and a similar bulk modulus has only been found experimentally in the low-pressure phase of monazite-type LaVO$_4$ and CeVO$_4$.[12,14] As $B_0$ for the monoclinic phase of PrVO4 has been calculated here taking in account also the cell parameters at pressure release (i.e. near ambient pressure), it is reasonable to compare it to that of the low-pressure phase of monoclinic LaVO$_4$. The reduced symmetry of the monoclinic phase allows for a large flexibility of the PrO$_9$ units and a structural distortion of VO$_4$ tetrahedral units which is forbidden in high-symmetry zircon structures.[50] As such, the VO$_4$ units (and not only PrO$_9$ units) for monoclinic *RE*VO$_4$ also contribute to the volume reduction of the unit cell, unlike to what happens in zircon or scheelite. Hence, the low bulk modulus of the high-pressure phases of PrVO$_4$ observed here might be thus a consequence of the more efficient packing of the monoclinic structure and the flexibility of PrO$_9$ and VO$_4$ polyhedra.[22] Additionally, the high compressibility of the first HP phase of PrVO$_4$ may be related to the broadening of the Bragg peaks after the transition, which affects the accuracy of the unit-cell parameters. The broadening, in turn, is caused by the stress due to the large volume collapse at the transition. A further volume collapse, of about 9.18%, is also detected at the monazite-BaWO$_4$-II transition. This variation comes with a change in the coordination of Pr (from 9 to 10) and V (from 4 to 6), which increases the packing of the structure.[45] As a consequence, the post-monazite phase has a high bulk modulus of 147(6) GPa. These experimental bulk moduli compare favorably with the values obtained by fitting the *P-V* data from the *ab initio* simulations (Figure 9), taking in account the inherent overestimation of the initial volume given by GGA-PBE. As a very remarkable result, theoretical simulations also predict a higher compressibility for the monazite phase with respect to zircon, in perfect agreement with our experimental results (Table 2). It is interesting to note that the experimental value of $B_0$ for the BaWO$_4$-II phase value is in full agreement with the calculated bulk modulus of the post-monazite phase of LaVO$_4$, and the transition in both compounds occurs with a similar volume collapse (around 8%).[45] This finding suggests that the BaWO$_4$-II structure may be the post-monazite phase also for other orthovanadates. For instance, BaWO$_4$-II has been proposed as the post-monazite phase for NdVO$_4$ on the basis of theoretical simulations, with a similar calculated value for the bulk modulus.[51] In this respect, it is well known that the transition sequence in *RE*VO$_4$ vanadates is determined, in the first instance, by the size of the *RE* trivalent cation. NdVO$_4$ and vanadates with larger cations follow the zircon-monazite-post-monazite sequence,[51] while SmVO$_4$ and vanadates with smaller cation radii are expected to undergo a zircon-scheelite-fergusonite sequence.[52] Sm and Nd are thus considered the boundary that determines the behavior of the *RE*VO$_4$ compounds under high-pressure. On the other hand, in the proximity of the Sm-Nd limit, or for compounds with a RE cation with an *effective* size halfway between Nd and Sm (mixed vanadates), the equilibrium is unstable and may be shifted due either to experimental conditions or to energetic reasons. For instance, when NdVO$_4$ is submitted to *hydrostatic* compression it follows the zircon-scheelite-fergusonite sequence, while mixed vanadates such as Sm$_{0.5}$Nd$_{0.5}$VO$_4$ follow an even more peculiar path, with scheelite and monazite coexistence in a limited pressure range. The ionic radii of Pr$^{3+}$ and Nd$^{3+}$ is similar (0.99 vs. 0.983 Å respectively[53]), hence it is reasonable to expect that the behavior of PrVO$_4$ may change with different experimental conditions. For reference, a plot with the different phases observed for *RE*VO$_4$ compounds as a function of pressure and cation size is shown in Figure 10, with the different stability domains clearly delineated for each phase. It must be mentioned that in the plots of Figures 7, 8 and 9 a data point at 12.7 GPa was omitted, as the coexistence of the two monoclinic structures and the broadening of the Bragg peaks prevented the obtention of reliable cell parameters.

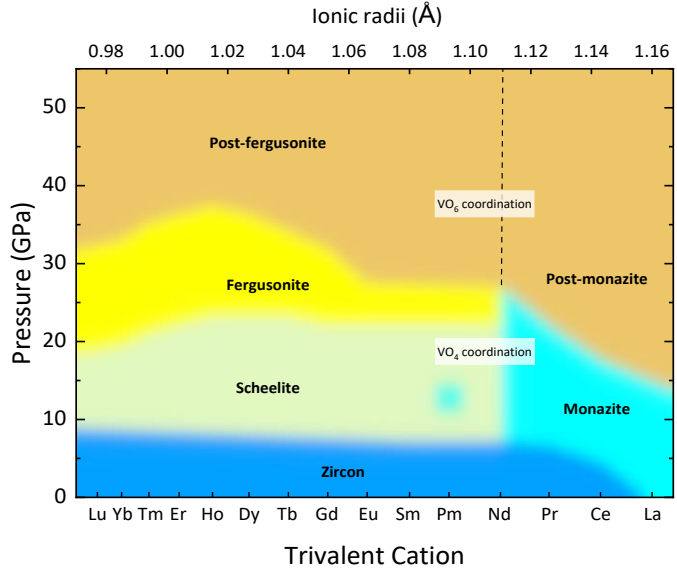

**Figure 10:** Phase diagram of *RE*VO$_4$ compounds as a function of element (equivalently, cation size) and pressure. The cyan gradient in correspondence of Pm corresponds to temporary appearance of the minority monazite phase detected in Sm$_{0.5}$Nd$_{0.5}$VO$_4$ from 9.2 to 13.6 GPa (Ref.22)

| Compound | Structure | PTM | $V_0$ (Å$^3$) | $B_0$ (GPa) | $B'_0$ | Reference |
|---|---|---|---|---|---|---|
| PrVO$_4$ | Zircon | MEW | 350.4(3) | 120(3) | 4 | This work |
| | Zircon | Theory | 365.05(15) | 119(2) | 4 | This work |
| | Monazite | MEW | 321(4) | 95(6) | 4 | This work |
| | Monazite | Theory | 336.7(2) | 101(1) | 4 | This work |
| | BaWO$_4$-II | MEW | 561(4) | 147(6) | 4 | This work |
| | BaWO$_4$-II | Theory | 608.0(4) | 125(2) | 4 | This Work |
| LaVO$_4$ | Zircon* | ME | 364.02 | 93(2) | 4 | 54 |
| | Monazite | MEW | 333.2(1) | 106(1) | 4 | 14 |
| | Monazite | Theory | 334.2(5) | 99(5) | 4 | 14 |
| | Monazite | Theory | 328.2 | 105.2 | 4.3 | 45 |
| | BaWO$_4$-II | Theory | 609.2 | 154 | 4.2 | 45 |
| CeVO$_4$ | Zircon | ME | 356.53 | 112(3) | 4 | 50 |
| | Zircon | Ne | 352.9 | 125(9) | 4 | 21 |
| | Zircon | Me | 354.86 | 118.9 | 4 | 55 |
| | Monazite | ME | 328.23 | 98(3) | 4 | 12 |
| | Monazite | Ne | 326.2(8) | 133(5) | 4.4(6) | 21 |
| | Monazite | Me | 327.21 | 142 | 4.4 | 45 |
| TbVO$_4$ | Zircon | Ne | 324.4(9) | 122(5) | 6.2(5) | 21 |
| TmVO$_4$ | Zircon | He | 312.6(1) | 120(2) | 4 | 47 |

**Table 2.** EOS of PrVO$_4$ (2$^{nd}$ order Birch-Murnaghan for all phases, in this work) and comparison with other *RE*VO$_4$ in zircon, monazite and BaWO$_4$-II phase. $V_0$, volume at ambient pressure, $B_0$ bulk modulus, $B_0'$ first derivative of bulk modulus. The asterisk in zircon LaVO$_4$ indicate that these values refers are measured on nanorods sample.

The discussion of the transformation pathway between the three structural phases of PrVO$_4$ is beyond the scope of this work. The most we can state here is that both transitions are reconstructive and involve the formation of new bonds. This can be seen in the Figure 11 where we compare the three structures. In order to facilitate the comparison, we have represented the zircon structure using the same space group and setting used for the monazite and post-monazite structures. This can be done since $P2_1/n$ is a subgroup of $I4_1/amd$. The group subgroup sequence is $I4_1/amd$ -> $Imma$ -> $Pnma$ -> $P2_1/n$. The structure obtained via this transformation should be rotated using the matrix $\begin{pmatrix} 0 & 0 & 1 \\ 1 & 0 & 0 \\ 0 & 1 & 0 \end{pmatrix}$ in order to describe zircon structure using the same space group and setting that it is used for the HP structures. In Figure 11 we can observe that, in spite of the possibility of describing zircon with the same space group as monazite and the similitudes between the edge-sharing V and Pr polyhedral chains in both structures, the monazite structure cannot be obtained via simple displacements of atoms; the transition, in fact, involves not only the displacement and twisting of the polyhedral units of monazite relative to analogous polyhedral units in zircon, but also the formation of a new Pr-O bond transforming PrO$_8$ into PrO$_9$.[56] In the case of the second transition, the changes of polyhedral units are more drastic. The displacement and tilting of VO$_4$ tetrahedra and PrO$_9$ polyhedra as a consequence of the transition leads to a change of the chemical environment of V and Pr, creating space for the construction of VO$_6$ distorted octahedra from VO$_4$ and PrO$_{10}$ polyhedra from PrO$_9$.[57]

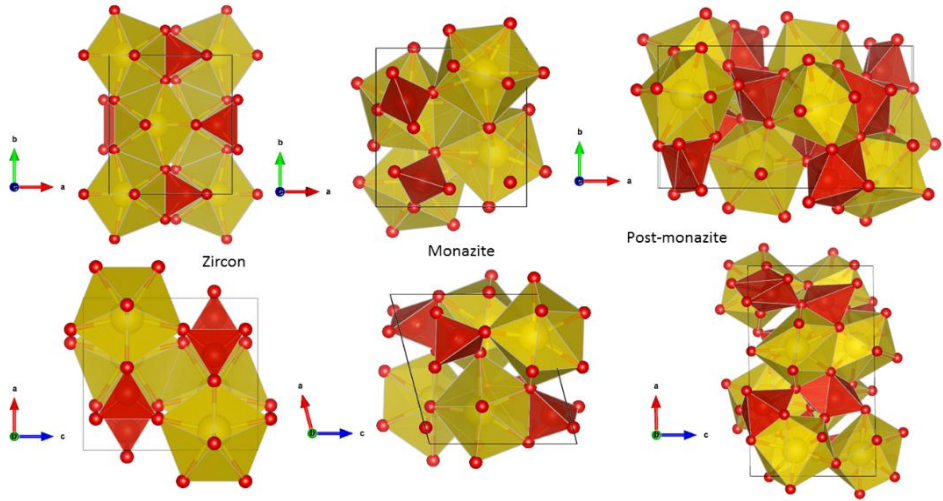

**Figure 11:** comparison of the zircon, the monazite and the BaWO$_4$ structure, shown using for all of them the same space group ($P2_1/n$).

We will now discuss the influence of pressure on the electronic properties of PrVO$_4$. In Fig. 12 we present the results of optical-absorption measurements. As we may observe in the figures, PrVO$_4$ presents a wide band gap, which can be assigned to a direct transition as in other zircon-type vanadates.[58] A value of 3.83(2) eV is obtained for the band-gap energy when a Tauc analysis is applied (Figure S1).[59] This approach is known to work well for oxides of this family.[60] The obtained band-gap energy is very similar to that of other zircon-type vanadates (for comparison, see also Table 3, which reports the value of the band-gap energy for different orthovanadates).[60] For zircon PrVO$_4$ DFT simulations predict a direct band gap at the high symmetry Γ point with a value of 3.045 eV for $E_g$ (See Figure S2). The underestimation of the band-gap of zircon-phase orthovanadates by DFT calculations is not an unusual result.[45,47] The inclusion of the Hubbard $U$-potential increases the magnitude of the gap, compared to when only PBE is applied. However, and since $U$ is an empirical parameter, we must note the band-gap width is dependent on this parameter. On a further increase of $U$, the magnitude of the gap could increase, albeit the lattice parameters are affected, increasing their magnitude. Nonetheless, DFT+$U$ can treat structural relaxations adequately and at the same time retain the correct physics of the on-site correlation for the V $d$-states. Moreover, DFT simulations of electronic structure are useful to explain the behavior of the band gap under high-pressure conditions. As can be seen in Figure 12, up to 5.3 GPa the absorption edge (and consequently the band-gap energy) of $PrVO_4$ is weakly affected by pressure. However, in subsequent compression steps we found an abrupt red shift of the band gap. This change can be correlated with the zircon-monazite transition detected by XRD at a similar pressure. A similar behavior has been observed at the same transition in $NdVO_4$. Upon further compression, up to 11.4 GPa, the absorption edge slightly redshifts. Beyond this pressure value several cracks started to develop in the crystal, which in case of further developing would preclude the performance of absorption measurements. Therefore, we decided to release the pressure. We think the cracks are related to precursor defects of the second phase transition observed at a higher pressure. Under decompression we observed that the phase transition is not reversible.

| Compound | Structure | $E_g$ (eV) | Reference |
|---|---|---|---|
| $PrVO_4$ | zircon | 3.83(2) | This work |
| $LuVO_4$ | zircon | 3.76 | 58 |
| $LaVO_4$ | monazite | 3.5(2) | 61 |
| $CeVO_4$ | zircon | 3.36 | 62 |
| $NdVO_4$ | zircon | 3.72 | 11 |
| $GdVO_4$ | zircon | 3.87 | 63 |
| $SmVO_4$ | zircon | 2.28 | 2 |
| $YVO4$ | zircon | 3.78 | 58 |
| $YbVO4$ | zircon | 3.79 | 58 |

**Table 3:** Experimental values of the band gap ($E_g$) for different orthovanadate compounds.

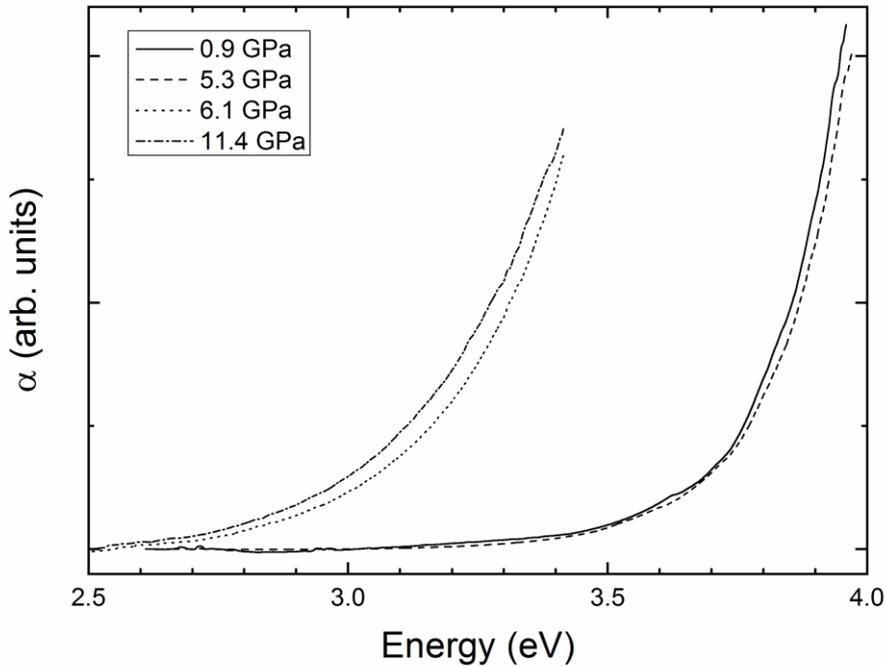

**Figure 12.** Absorption spectra of $PrVO_4$ at different pressures.

From our experiment we obtained the pressure dependence of the band-gap energy (Figure 13). In the low-pressure phase, the band gap opens with a pressure coefficient of $dE_g/dP = 2.6(6)$ meV/GPa. This behavior is similar to that of $NdVO_4$ and related vanadates.[60] The small increase of the band-gap with pressure is due to the decrease of the V-O distances. In fact, it is known that the top of the valence band and the bottom of the conduction band in zircon-type vanadates are dominated by V 3d and O 2p states. Our DFT simulations support this in the case of zircon $PrVO_4$ (Figure S2). DFT simulations also predict a positive pressure coefficient for the band gap of the pressure phase, $dE_g/dP = 11.8(12)$ meV/GPa, in agreement with experiments. Akin to what has been observed in other zircon vanadates, the shrinking of V-O bonds with pressure enhances the repulsion between bonding and anti-bonding states, causing the opening of the gap. The $VO_4$ tetrahedron is known to be hardly compressible.[13] This is the reason of the weak variation of the gap with pressure for this low-pressure phase. As observed in NdVO$_4$, at the zircon-monazite transition there is an abrupt decrease of the band gap, whose energy value is 3.30(5) eV at 6.1 GPa. This collapse is caused by the atomic reorganization after the phase transition, which lead to an enhancement of orbital hybridization. Our calculations assign to the monazite an indirect band gap, with the valence band maximum (VBM) and the conduction band minimum (CBM) still dominated by the V and O orbitals and located close to the A and the E point respectively (see Figure S3). The indirect character of the band gap and the contribution to it are maintained across the entire pressure range, along with the position of the VBM and the CBM (Figure S4). In the HP phase the band gap closes with a pressure coefficient of -13.7(5) meV/GPa. This is a consequence of the enhancement of orbital hybridization as pressure increases. DFT correctly estimates the band gap of the monazite phase, despite it predicting a small positive pressure coefficient for $E_g$, in contrast with experimental evidence. This divergence between theory and experiments has been observed in other vanadates and has been attributed to different conditions for the simulations and the experiments regarding temperature and hydrostaticity. The latter may be important in the specific case of monazite-type PrVO$_4$, since we are in proximity of the hydrostatic limit of the PTM.[16] Also, the presence of excitonic effects may play a role: at room temperature, broad excitonic peaks may overlap with the absorption tail of monazite and decrease its steepness (as observed in Figure 12), thus masking an increase of the band gap.[58] As mentioned before, the band gap of the post-monazite phase could not be measured experimentally, due to cracks in the samples in proximity of the second transitions. Hence, for completeness we have calculated the band gap of the BaWO$_4$-II phase and its evolution with pressure. According to DFT simulations, at 12.2 GPa, PrVO$_4$ is an indirect semiconductor with a band gap of around 2.5 eV. The valence and the conduction band are still dominated by the oxygen and the vanadium orbitals, respectively, with the VBM and CBM located at the Γ and the E points (see Figure S5). Thus, a band gap collapse is expected at the second transition, in agreement with the change on coordination of the V-O units that we have described before. According to DFT, the gap is expected to open under compression, reasonably due to the shrinking V-O bonds. These theoretical findings support that metallization is not happening in our sample. This is consistent with the fact that the sample does not became black after the second transition.

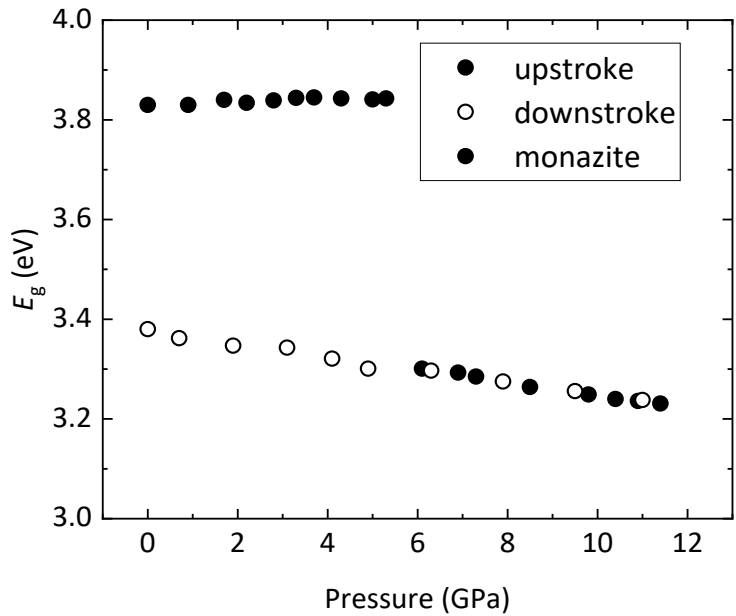

**Figure 13:** Experimental pressure dependence of the band-gap energy. Black symbols correspond to compression experiments and white symbols to decompression experiments.

To conclude, we will present the results of electrical resistivity measurements. As can be seen in Figure 14, the resistivity of the low-pressure phase presents a slight decrease as a function of pressure. $PrVO_4$ is known to be a high-resistivity material, in which electron conduction is mediated by the hopping mechanism between f-orbitals.[64] These are known to gradually delocalize under compression leading thus to the observed decrease of resistivity.[65] Between 5.3 and 6 GPa we observed a steep decrease of the resistivity, of about one order of magnitude. This is consistent with the occurrence of the zircon-monazite transition which triggers a collapse of the band gap. Across the entire pressure range of the experiments (which is limited by the experimental set-up) we have observed a smooth decrease of the resistivity with pressure, without any indication of the second phase transition. Under pressure decrease the change in the resistivity is not reversible, in agreement with what is found in XRD and optical experiments. The large decrease of the resistivity at the phase transition can have a twofold origin: the decrease of the band gap detected by absorption experiments and the formation of defects generated during the first-order phase transition from zircon to monazite structure; after the occurrence of the phase transition the defects could behave as donor states.[66,44] Lets discuss the first possibility. The phase transition involves a 15% change of the band-gap energy. This alone cannot explain a decrease of resistivity of more than one order the magnitude as observed in our experiment (see Fig. 14). Other phenomena, like an increase of the extrinsic free-carrier concentration, associated probably to the creation of donor states (related to defects), or an enhancement of the free-carrier mobility, should also contribute to the abrupt decrease of the resistivity at the phase transition. According to calculations, the dispersion of the conduction band is qualitatively similar in the zircon and monazite structures; thus, the effective mass of electrons will not be very sensitive to the phase transition, not affecting substantially the mobility. On the other hand, the creation of defects (associated to the large volume collapse of the transition) could only trigger a decrease of the electron mobility (which will increase resistivity), by enhancing the scattering of electrons with defects. Therefore, the contribution of carrier mobility can be neglected, having probably the creation of donors, related to defects, the most important contribution to the observed resistivity decrease. Studies combining pressure and temperature should be carried out to give a definitive answer to the observed decrease of resistivity.[67] It is important to highlight here that no evidence of pressure-induced metallization has been found here, both in optical and resistivity studies.

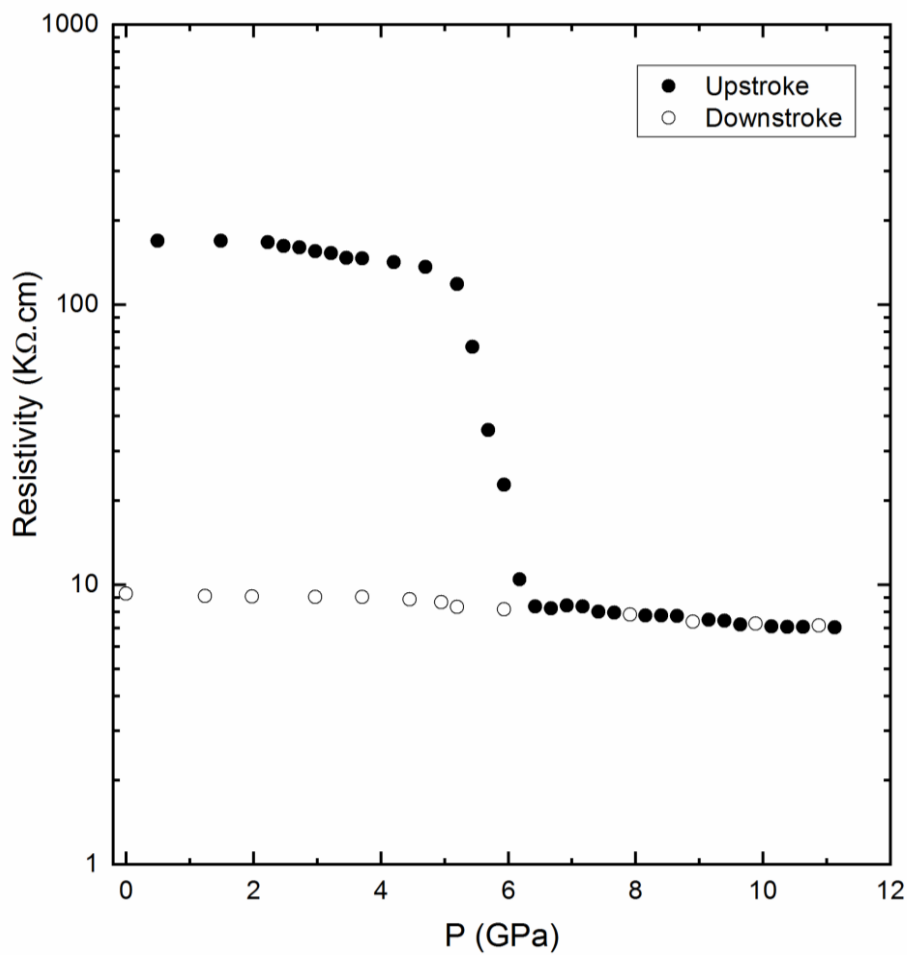

**Figure 14:** resistivity of PrVO$_4$ versus pressure. Black symbols correspond to compression and white symbols to decompression.

## Summary

We have performed a study of PrVO$_4$ orthovanadate under high-pressure conditions by means of synchrotron powder x-ray diffraction, optical-absorption, and resistivity measurements, and complemented them with DFT simulations at high-pressure conditions. We confirmed that PrVO$_4$ undergoes a first-order transition around 6.2 GPa to a monoclinic monazite-type structure, with a volume collapse of about 9.6%. Further, a second transition to a monoclinic BaWO$_4$-II structure is found around 14.5 GPa, again with a sharp decrease in the volume. Nonetheless, the monazite phase is recovered upon pressure releases. We suggest that the BaWO$_4$-II structure can occur as the post-monazite phase of other orthovanadates, by comparison with previous results obtained on studies of NdVO$_4$ and LaVO$_4$. Both transitions are correctly predicted by theoretical enthalpy calculations. The compressibility of the three structures is found to be anisotropic. The EOSs of all phases have also been calculated. The bulk modulus of the zircon phase is found to be comparable to that of soft rare-earth vanadates, such as CeVO$_4$, TbVO$_4$ and TmVO$_4$. Finally, the zircon-monazite transition strongly modifies the electronic properties of PrVO$_4$. Both the band-gap energy and resistivity show and abrupt decrease at the transition point. Reasons for these phenomena are discussed, being linked to an enhancement of orbital hybridization induced by pressure. DFT simulations confirmed that zircon PrVO$_4$ is a direct semiconductor, unlike its monazite and post-monazite phase. For all the structural phases, the band gap is dominated by the levels of V and O atoms. We have also calculated a band gap collapse in correspondence of the second transition, which could not be observed experimentally. This collapse may be explained by the atomic reorganization at the transition and the increase in the coordination of the vanadium atom following it. The overall DFT simulations support our conclusions. The divergences that were found can be explained within the limitations of the DFT calculations and their assumptions, with respect to the experimental conditions.

## Associated Content

The Supporting Information is available free of charge at XXXXXXXXXX.

Axial compressibility for the zircon, monazite and $BaWO_4$-II -type phases; powder XRD measurements, details of the data collections, refinement results; Tauc plots of the optical absorption experiments; structural data obtained for $PrVO_4$ at different pressures; band diagrams for zircon, monazite and $BaWO_4$-type $PrVO_4$.

## Author Contributions

The manuscript was written through contributions of all authors. All authors have given approval to the final version of the manuscript

## Notes

The authors declare no competing financial interests

## Acknowledgments

Research was supported by the Spanish Ministerio de Ciencia, Innovación y Universidades, the Spanish Research Agency, and the European Fund for Regional Development under Grant Nos: MAT2016-75586-C4-1-P, PID2019-106383GB-C41 and RED2018-102612-T and by Generalitat Valenciana under Grant Prometeo/2018/123 (EFIMAT). C.P. and J. A. S. thank the financial support from Spanish Ministerio de Economia y Competividad through FIS2017-83295-P project. J. A. S. wants also to thank the financial support of the Ramón y Cajal fellowship (RYC-2015-17482). Powder x-ray diffraction experiments were performed at the Materials Science and Powder Diffraction beamline of ALBA Synchrotron. We acknowledge the computer resources of the ARCHER UK National Supercomputing Service via membership of the UK's MCC consortium, where calculations were carried out. E.L.d.S would like to thank financial support of the project Ref. NORTE-01-0145-FEDER-022096, Network of Extreme Conditions Laboratories (NECL), financed by FCT and co-financed by NORTE 2020, through the program Portugal 2020 and FEDER.

Sn Related Donor Impurities. *Semicond. Sci. Technol.* **2003**, *18*, 241–246.

Supporting Information

# PrVO$_4$ under High Pressure: Effects on Structural, Optical and Electrical Properties

*Enrico Bandiello*[1,*], *Catalin Popescu*[2,*], *Estelina Lora da Silva*[3,4,] *Juan Ángel Sans*[4], *Daniel Errandonea*[1], *Marco Bettinelli*[5]

[1]Departamento de Física Aplicada-ICMUV, MALTA Consolider Team, Universidad de Valencia, Edificio de Investigación, C/Dr. Moliner 50, Burjassot, 46100 Valencia, Spain

[2]CELLS-ALBA Synchrotron Light Facility, Cerdanyola del Valles, 08290 Barcelona, Spain

[3]IFIMUP, Departamento de Física e Astronomia, Faculdade de Ciencias da Universidade do Porto, Porto, Portugal

[4]Instituto de Diseño para la Fabricación y Producción Automatizada, MALTA Consolider Team, Universitat Politècnica de València, 46022 València, Spain

[5]Luminescent Materials Laboratory, Department of Biotechnology, University of Verona and INSTM, UdR Verona, Strada Le Grazie 15, 37134 Verona, Italy

[*]Corresponding authors: Enrico Bandiello; email: enrico.bandiello@uv.es. Catalin Popescu: cpopescu@cells.es

**Table S1.** Linear compressibility for zircon PRVO$_4$, as obtained by powder XRD data (Experiment) and DFT data (DFT).

|  |  | zircon |
|---|---|---|
| $k_a$ (10$^{-3}$ GPa$^{-1}$) | Experiment | 3.1(1) |
|  | DFT | 2.76(4) |
| $k_b$ (10$^{-3}$ GPa$^{-1}$) | Experiment | 3.1(1) |
|  | DFT | 2.76(4) |
| $k_c$ (10$^{-3}$ GPa$^{-1}$) | Experiment | 1.29(8) |
|  | DFT | 2.12(3) |

**Table S2.** Compressibility tensor for monazite and BaWO$_4$-II-type PRVO$_4$ ($\beta_{ij}$), as obtained by powder XRD data as obtained by powder XRD data (Experiment) and DFT data (DFT). Eigenvectors and eigenvalues are also reported ($\lambda_i$ and $ev_i$, respectively).

|  | monazite (6.2 GPa) | | BaWO4-II (14.4 GPa) | |
|---|---|---|---|---|
|  | **Experiment** | **Theory** | **Experiment** | **Theory** |
| $\beta_{11}$ (10$^{-3}$ GPa$^{-1}$) | 3.258 | 2.893 | 1.079 | 1.915 |
| $\beta_{22}$ (10$^{-3}$ GPa$^{-1}$) | 4.352 | 2.749 | 1.640 | 1.570 |
| $\beta_{33}$ (10$^{-3}$ GPa$^{-1}$) | 2.394 | 2.405 | 1.729 | 1.645 |
| $\beta_{13}$ (10$^{-3}$ GPa$^{-1}$) | -1.182 | -0.453 | 0.164 | -0.612 |
| $\lambda_1$ (10$^{-3}$ GPa$^{-1}$) | 3.824 | 3.163 | 1.040 | 2.047 |
| $ev_1$ | (-0.873, 0.000, 0.488) | (-0.859, 0.000, 0.513) | (-0.973, 0.000, 0.232) | (-0.795, 0.000, 0.626) |
| $\lambda_2$ (10$^{-3}$ GPa$^{-1}$) | 4.484 | 2.749 | 1.640 | 1.570 |
| $ev_2$ | (0.000, 1.000, 0.000) | (0.000, 1.000, 0.000) | (0.000, 1.000, 0.000) | (0.000, 1.000, 0.000) |
| $\lambda_3$ (10$^{-3}$ GPa$^{-1}$) | 1.998 | 2.135 | 1.764 | 1.115 |
| $ev_3$ | -(0.488, 0.000, 0.873) | -(0.513, 0.000, 0.859) | -(0.232, 0.000, 0.973) | -(0.626, 0.000, 0.779) |

**Table S3.** Equilibrium parameters for zircon, monazite, and BaWO4-II phase as calculated by DFT.

| | Zircon | Monazite | BaWO$_4$-II |
|---|---|---|---|
| Pressure | 1 bar | 1 bar | 14.93 GPa |
| $a_0$ (A) | 7.47 | 7.08 | 12.45 |
| $b_0$ (A) | 7.47 | 7.30 | 6.51 |
| $c_0$ (A) | 6.54 | 6.75 | 6.85 |
| β (degree) | - | 105.34 | 96.20 |
| $V_0$ (A$^3$) | 364.54 | 336.63 | 552.01 |
| $B_0$[GPa] | 111.82 | 102.40 | 122.21 |
| $B_0´$ | 5.1 | 3.80 | 4.17 |

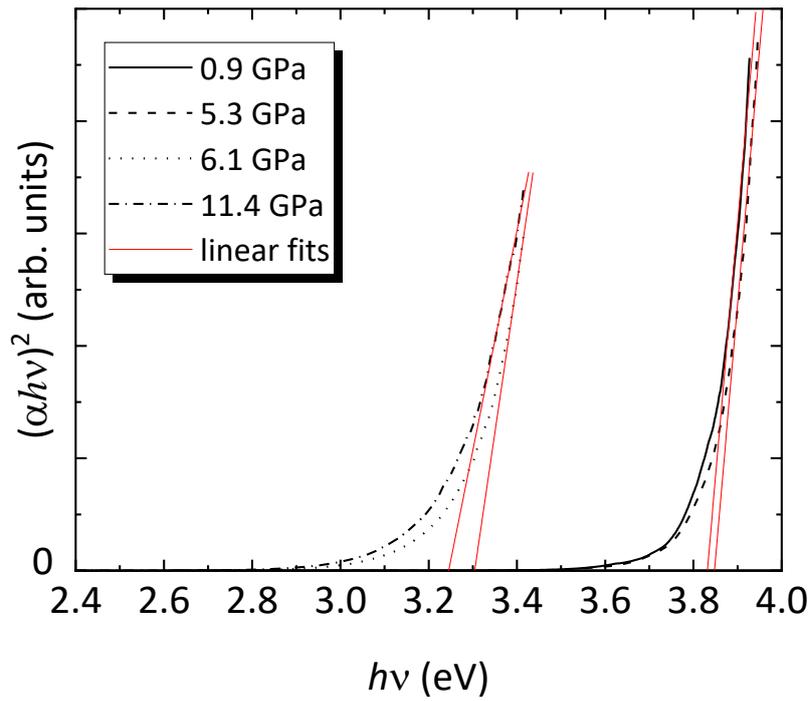

**Figure S1**. Tauc plot of the absorption spectra of PrVO$_4$ reported in the main text, with the linear fits extrapolated onto the x-axis for the determination of the band gap (red lines).

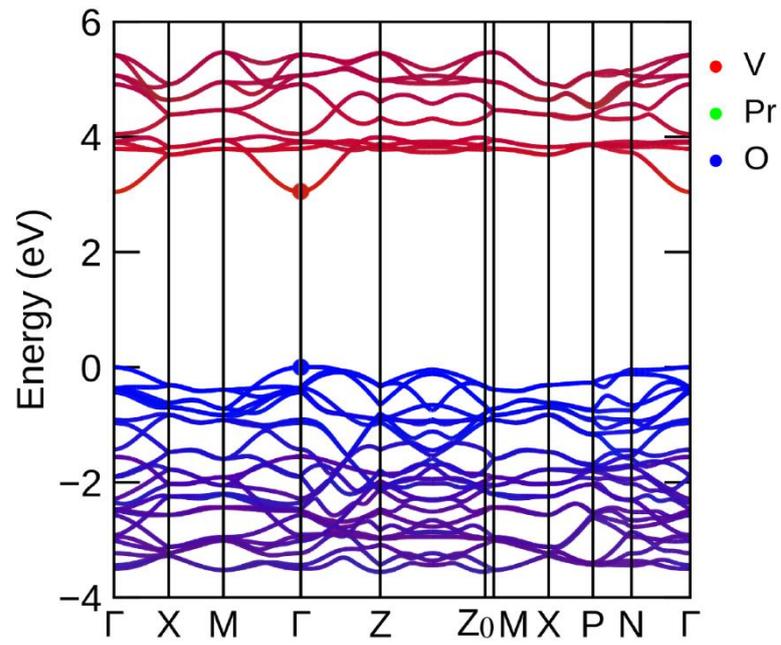

**Figure S2.** Electronic band dispersion of the zircon phase at ambient pressure. The red circle corresponds to the conduction band minima and the blue circle to the valence band maximum (located at $\Gamma$ point), being the system a direct gap insulator with a gap width of 3.0 eV.

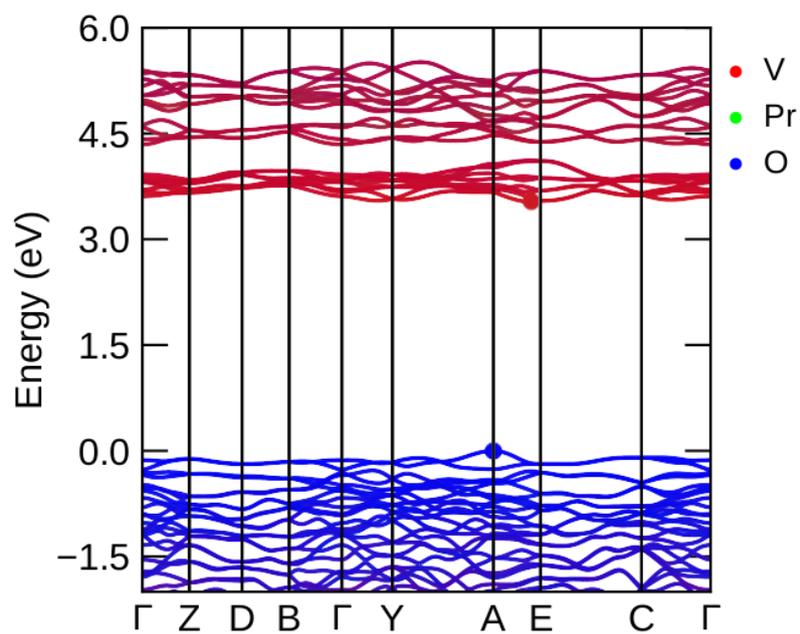

**Figure S3.** Electronic band dispersion of the monazite phase at 7 GPa. The red circle corresponds to the conduction band minima (A point) and the blue circle to the valence band máximum (in the proximity of the E point), being the system an indirect gap insulator with a gap width of 3.53 eV.

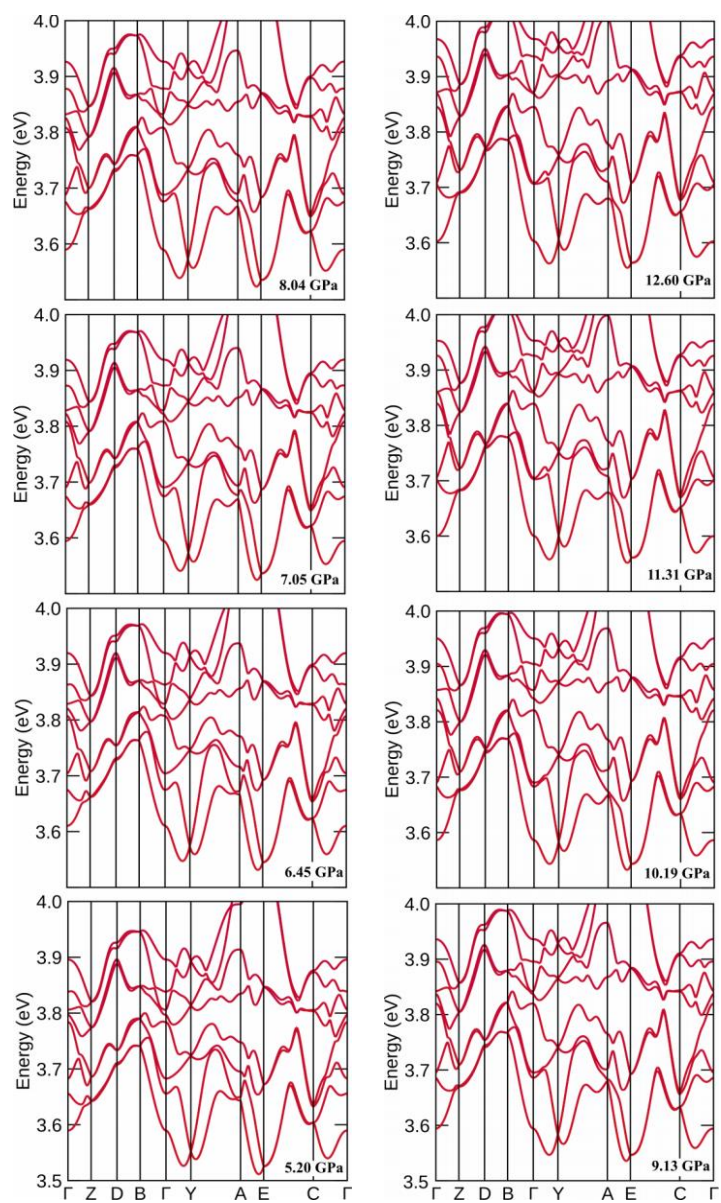

**Figure S4**. Electronic band dispersion for the conduction band of monazite-type PrVO$_4$ at selected pressures. The conduction band minimum is consistently located close to the A point in the entire studied pressure range.

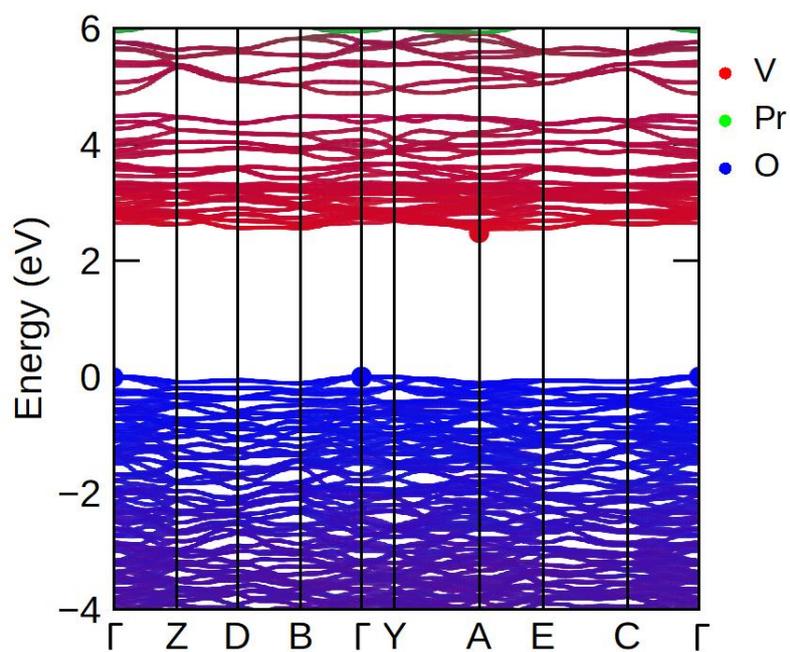

**Figure S5.** Electronic band dispersion of BaWO$_4$-II-type PrVO$_4$ at 12.2 GPa. . The red circle corresponds to the conduction band minima (A point) and the blue circle to the valence band máximum (Γ point), being the system an indirect gap insulator with a gap width of 2.5 eV

**Table S4:** Details of the data collections, refinement results, and structural data as obtained by powder-XRD experiments on PrVO4 for the zircon, the monazite and the BaWO4-II phase. The pressure at which the refinements were performed is indicated in the table. For the monazite phase, the refinement was performed at 0.4 GPa, after pressure release. In the definition of $R$ factors $Y_i^{obs}, Y_i^{calc}$ indicate observed and calculated peak intensity values, respectively; $w_i$ are the weights and $N$ is the number of data points.

| | Zircon PrVO4 | Monazite PrVO4 | BaWO4-II PrVO4 |
|---|---|---|---|
| Source | synchrotron | synchrotron | synchrotron |
| Chemical Formula | PrVO4 | PrVO4 | PrVO4 |
| Formula Weight | 255.85 | 255.85 | 255.85 |
| $T$/K | 298 | 298 | 298 |
| Pressure | $10^{-4}$ GPa | 0.4 GPa | 14.45 GPa |
| Wavelength/Å | 0.4246 | 0.4246 | 0.4246 |
| Crystal System | tetragonal | monoclinic | monoclinic |
| Space Group | 141 | 14 | 14 |
| $a$/Å | 7.3625(2) | 6.9930(8) | 11.557(7) |
| $b$/Å | 7.3625(2) | 7.1796(7) | 6.522(7) |
| $c$/Å | 6.4614(4) | 6.6311(8) | 6.847(7) |
| $\alpha$/deg | 90 | 90 | 90 |
| $\beta$/deg | 90 | 104.60(1) | 90.71(1) |
| $\gamma$/deg | 90 | 90 | 90 |
| $V$/Å$^3$ | 350.26(2) | 322.2(8) | 516.1(8) |
| $Z$ | 4 | 4 | 8 |
| d-space range/Å | 6.08-1.29 | 6.08-1.29 | 6.08-1.29 |
| $\chi^2$ | 5.6169 | 2.4458 | 2.9241 |
| $R_p$ | 0.3719 | 0.3325 | 0.2476 |
| $R_{wp}$ | 0.5259 | 0.5200 | 0.4261 |
| Definition of $R$ factors | $R_p = \Sigma_i \dfrac{\left|Y_i^{obs} - Y_i^{calc}\right|}{\Sigma_i Y_i^{obs}}$  $R_{wp}^2 = \Sigma_i \dfrac{w_i\left(Y_i^{obs} - Y_i^{calc}\right)^2}{\Sigma_i^n w_i\left(Y_i^{obs}\right)^2}$  $R_{exp}^2 = \dfrac{N}{\Sigma_i^n w_i\left(Y_i^{obs}\right)^2}$  $\chi^2 = \left(\dfrac{R_{wp}}{R_{exp}}\right)^2$ | | |